\documentclass[%
reprint,
 amsmath,amssymb,
 aps, graphicx,
longbibliography,
floatfix,
]{revtex4-1}

\usepackage{xcolor}
\usepackage{graphicx}% Include figure files
\usepackage{dcolumn}% Align table columns on decimal point
\usepackage{bm}% bold math
\begin{document}

\preprint{APS/123-QED}

\title{Discrimination emerging through spontaneous symmetry breaking in a spatial prisoner's dilemma model with multiple labels}

\author{Gorm Gruner Jensen}
\email{ggjensen@itp.uni-bremen.de}
\author{Frederik Tischel}
\author{Stefan Bornholdt}
\affiliation{%
Institute for Theoretical Physics, University of Bremen, 28359 Bremen, Germany
}

\date{\today}

\begin{abstract}
Social discrimination seems to be a persistent phenomenon in many cultures.
It is important to understand the mechanisms that lead people to judge others by the group to which they belong, rather than individual qualities.
It was recently shown that evolutionary (imitation) dynamics can lead to a hierarchical discrimination between agents marked with observable, but otherwise meaningless, labels.
These findings suggest that it can give useful insight, to describe the phenomenon of social discrimination in terms of spontaneous symmetry breaking.
The investigations so far have, however, only considered binary labels.
In this contribution we extend the investigations to models with up to seven different labels.
We find the features known from the binary label model remain remarkably robust when the number of labels is increased.
We also discover a new feature, namely that it is more likely for neighbours to have strategies which are similar, in the sense that they agree on how to act towards a subset of the labels.
\end{abstract}

\maketitle{}

\section{Introduction}
\label{introduction}
% # Discrimination introduction
% What is discrimination? 
Discrimination is often defined as treating individuals differently because of the groups to which they belong rather than individual traits (or qualities).
Most of the empirically observed discrimination can be explained by ingroup favouritism \cite{brewer1999,Efferson2008}. 
That is, people tend to act in favour of those who are similar to themselves at the expense of those who are different. 
An increasing body of research, however, is suggesting that discrimination cannot be explained by ingroup favouritism alone.
A number of experiments have shown, that many inter-group relations are  asymmetric in the sense that members of one of the groups show much less, some times even negative, ingroup favouritism than members of the other \cite{Ridgeway1998,Rudman2002,Jost2002,Jost2004,March2014,Proestakis2016}.
These findings seem to indicate that there exists a sort of hierarchy of social status between different groups.
In this paper we will explore a minimalistic evolutionary game theory model in which persistent hierarchical discrimination can emerge through spontaneous symmetry breaking.

Most of the evolutionary game theory literature has been a search for mechanisms which promote cooperative behaviour through evolutionary dynamics \cite{smith1974, Nowak2006}.
One mechanisms which have been thoroughly studied is the so called tag-based cooperation \cite{Riolo2001, traulsen2003, Traulsen2007, Cohen2012, Garcia2014, Hadzibeganovic2014, Zhang2015}.
This is of special interest to us, because the introduction of observable tags makes it possible to define discrimination in a very simple way.
One can say that an agent's behaviour -- or strategy -- is discriminating if it is different towards peers who have different tags but identical behaviours.
Most of the models of tag-based cooperation are, however, not directly applicable for describing persistent hierarchical discrimination.
Some papers have already been very explicit about the close relation between tag-based cooperation and ingroup favouritism \cite{Fu2012, Zhang2015}.
Also, the behaviour of the tag-based cooperation models tends to express cyclic or wave-like dynamics, with a constant renewal of the dominating tags and strategies known as the `chromodynamics of cooperation' \cite{Traulsen2007}.

Recently, it was demonstrated that persistent hierarchical discrimination can emerge if the evolutionary dynamics only works on strategies, but not on the labels \cite{jensen2019imitating}.
By starting from a spatially extended system -- another well know mechanism for promoting cooperation \cite{nowak1992, Ohtsuki2006, perc2013} -- it was shown that under high selection pressure the cooperation would partially fail in a way which breaks the symmetry between two groups of agents distinguished only by an otherwise meaningless label.
It was shown that the proposed dynamics consistently leads to a state where, dependent on parameters, either the minority or the majority is systematically favoured. 
Human societies, however, consist of more than two types of people.
Humans can for example have many different religions, countries or origin, eye colours, and so on.
It is, therefore, natural to ask whether the same mechanism can also lead to discrimination if there are more than two different labels in the system.

In this paper we will examine an extension of the hierarchical discrimination model \cite{jensen2019imitating} with up to seven different labels.
This leads to a richer set of possible model outcomes, as the number of competing strategies grows exponentially with the number of labels.
We will show that much of the original structure is preserved, in particular that the number of labels have almost no impact on the parameter regions dominated by unconditional cooperation or defection. 
This is a remarkable result considering that the fraction of non-discriminating strategies decreases exponentially.

\section{Model}
\label{model}
Consider a population of $N$ agents occupying the nodes of a graph. The graph edges represent an agent's neighbourhood. 
Agents interact with their neighbours in a prisoner's dilemma type game where they can either cooperate or defect.
Each agent has one of $L$ distinct labels which can be observed by its neighbours. 
The label is the only observable difference between agents.  
An agent may therefore discriminate by cooperating with those neighbours who carry some of the labels while defecting against those carrying others.
A strategy specifies whether to cooperate with or defect against neighbours with each of the $L$ labels.
There are $2^L$ different possible strategies since a strategy can be represented by one binary variable for each label.

Cooperation costs an amount $c$ for the donating agent and gives a benefit $b$ to the receiver.
The payoff of an agent is calculated as the sum of the benefits the agent receives, minus the sum of the costs the agent pays:
\begin{align}
    p_i = \sum_{j\in \mathcal{N}_i} b\cdot C_j(\lambda_i) - c\cdot C_i(\lambda_j)
\end{align}
where $\mathcal{N}_i$ is the set of neighbours of agent $i$, $\lambda_i\in \{1,...,L\}$ is the label of agent $i$, and $C_i(\lambda_j)=1$ if it is the strategy of agent $i$ is to cooperate with the label of agent $j$ and $C_i(\lambda_j)=0$ if it is to defect. 
It should be noted that the payoff is a simple function of the state -- the labels and strategies of the agent and its neighbours -- and that it is not accumulated over time.

The dynamical variables in our model are the agents' strategies.
These change according to the following rule:
First we choose a random agent with uniform probability whose strategy will be updated.
With a small probability $\mu$ the agent will ``mutate'', i.e. choose a new strategy at random with uniform probability distribution.
Most of the times, with probability $1-\mu$, the selected agent will copy the strategy of one of its neighbours.
That neighbour is chosen with a probability proportional to its `fitness'.
The fitness of agent $i$, $f_i$, is related to its payoff $p_i$ via the expression $f_i = \exp(w\cdot p_i)$.
Here, $w$ is a global parameter which we will refer to as the selection pressure.
When the selection pressure is very small, $w\rightarrow0$, it is almost equally likely to choose any neighbour independent of their payoff.
When the selection pressure is large, $w\rightarrow\infty$, the neighbour with the largest payoff will almost certainly be chosen.

Notice that when an agent chooses a new strategy it is indifferent to how the strategy matches with the labels.
For example, there is nothing to hinder that an agent with a blue label copies a strategy from a neighbour with a green label which dictates only to cooperate with green neighbours.

In the case where there are $L=2$ different labels, our model is the same as that studied in \cite{jensen2019imitating}. 
In the case where there is only $L=1$ type of agents, it is a slight variation of a model presented in 2005 by Ohtsuki et al. \cite{Ohtsuki2006}, designed to demonstrate that evolutionary dynamics can promote cooperation in systems with spatial structure.
Our model varies from the one proposed by Ohtsuki et al. by using the exponential function ($f= \exp(w\cdot p)$) in the relation between fitness and payoff, rather than an affine function ($f=1-w+wp$).
The two functions converge in the limit of vanishing selection pressure ($w\rightarrow0$).

\section{Results}
\label{results}

\begin{figure}
    \centering
    \includegraphics{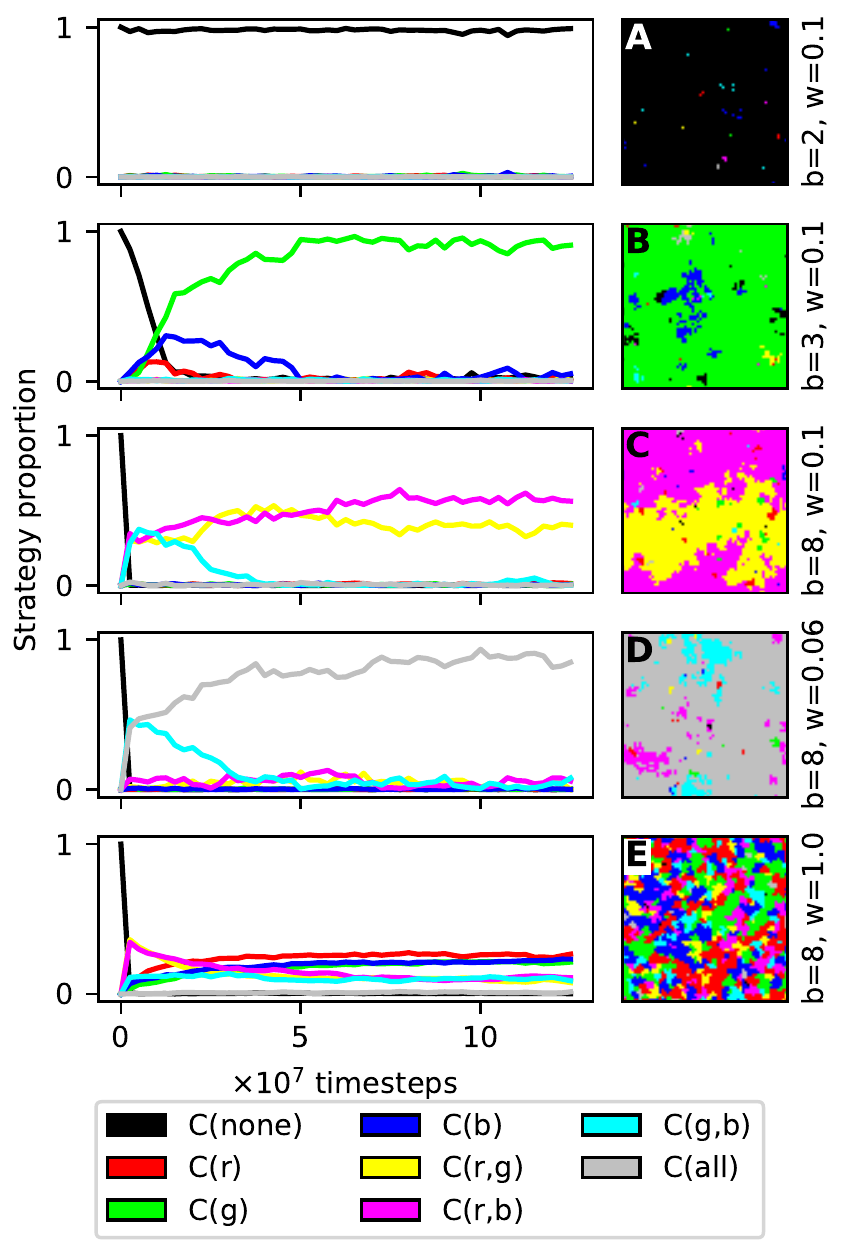}
    \caption{
    \textbf{Left:} Time-series of the strategy proportion for squared lattices with periodic boundary conditions, 10000 agents and 3 labels for different benefits \textit{b} and selection pressures \textit{w} with constant cost $c=1$ and constant mutation rate $\mu=0.001$. Starting with anyone having the strategy to not cooperate. \textbf{Right:} Snapshots of the strategy distributions corresponding to the end of the time-series. The chosen values of \textit{b} and \textit{w} correspond to the different phases shown in phase diagram in figure \ref{fig:paramscan_L=3}.
    \label{fig:examples}
    }
\end{figure}

\begin{figure*}
    \centering
    \includegraphics{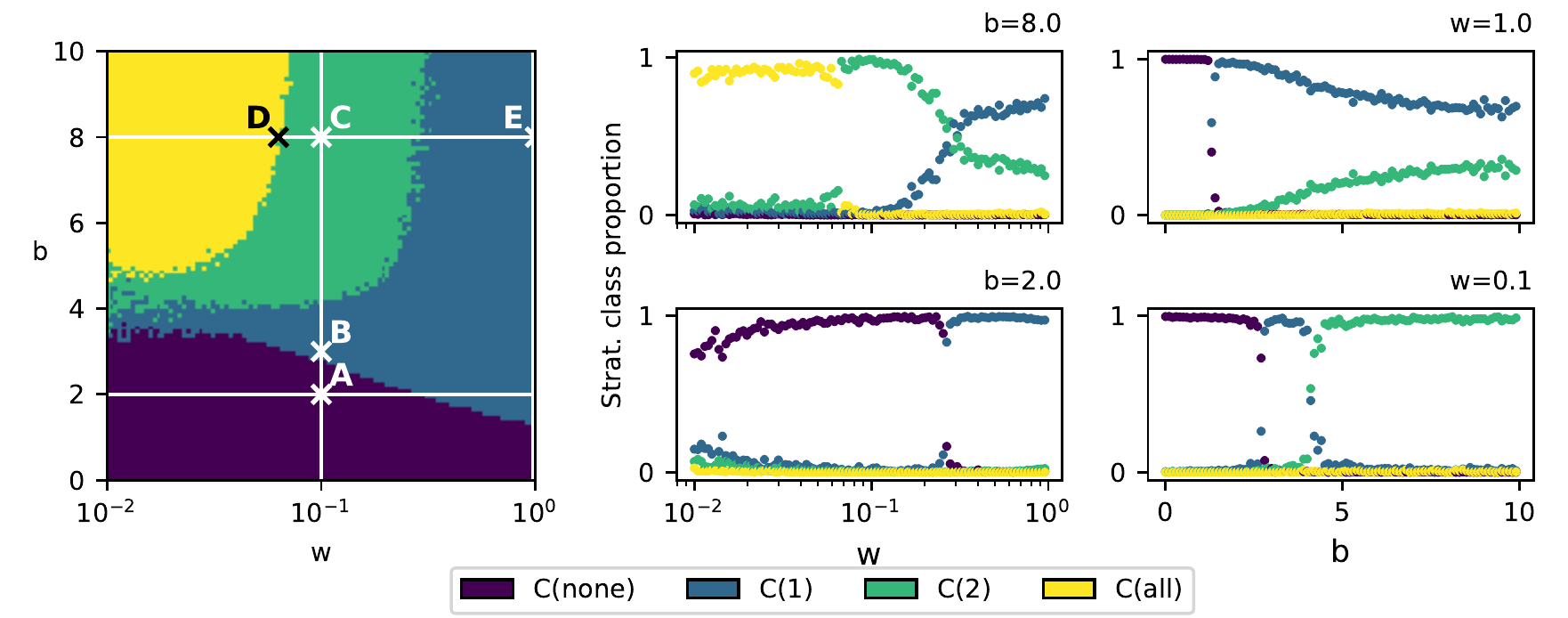}
    \caption{Results for squared lattices with periodic boundary conditions, 10000 agents and 3 labels, a constant mutation rate $\mu=0.001$ and a constant cost $c=1$. \textbf{Left:} phase diagram showing the most dominating strategy class over selection pressure \textit{w} and benefit \textit{b} after $1.25\times10^8$ time steps, the marks A-E correspond to the results shown in figure \ref{fig:examples}. \textbf{Middle:} Parameter scans showing strategy class proportion over selection pressure \textit{w} for fixed cooperation benefit \textit{b}, correspond to the horizontal lines in the phase diagram. \textbf{Right:} Parameter scans showing strategy class proportion over cooperation benefit \textit{b} for fixed selection pressure \textit{w}, correspond to the vertical lines in the phase diagram.}
    \label{fig:paramscan_L=3}
\end{figure*}

The main results presented in this paper are all obtained from a system with $100\times100$ agents arranged in a regular square lattice with periodic boundary conditions.
The labels are randomly assigned to each agent with independent uniform probability.
To reduce the number of parameters we have used a small constant mutation rate of $\mu=0.001$.
The non-zero mutation rate prevents the system from getting stuck in absorbing single-strategy fixed-points.
One benefit of this is that we can start every simulation in the simplest strategy-configuration in which no agent cooperates with anyone.
We have also explored the model with agents arranged in both random regular graphs and Erd\H{o}s-R\'{e}nyi graphs. 
These results are collected in section A of the supplementary material for comparison.

\subsection{Three labels}
To build an understanding of how our model behaves, we start by presenting five examples of typical dynamics arising when agents are arranged in a square lattice and the number of distinct labels is $L=3$ (see figure~\ref{fig:examples}). 
These examples illustrate the variety of behaviours observed at different values of cooperation benefit $b$ and selection pressure $w$.
Each example consists of two sub-figures: 
One time-series of the fraction of the population following each of the eight possible strategies, 
and one snapshot showing how the strategies are distributed on the lattice at the end of the time-series.
Because of the non-zero mutation rate it is impossible for the model to have any fixed-points, but the time-series clearly indicate that the systems tend to reach a meta-stable state in which the fractions of strategies fluctuates with small variations around some constant value.
In the four examples A--D the stationary distributions are clearly dominated by a single strategy. 
We find that this is a general result as long as the selection-pressure $w$ is small ($w \lesssim 0.2$).
This is reminiscent of the absorbing states in the voter model \cite{clifford1973, cox1986diffusive}.
When the selection-pressure is large, such as in example E, the situation is more complicated and requires a more careful analysis which we will return to later.

Each simulation was initiated with the strategy configuration in which all agents are `not cooperating with anyone'.
The parameters chosen for each of the five examples are marked in the phase diagram in figure~\ref{fig:paramscan_L=3}.

\textbf{A:} 
Our first example has a low cooperation benefit $b=2$ and intermediate selection pressure $w=0.1$. At these parameters, the system stays dominated by the strategy $C(\text{none})$ (not cooperating with anyone).

\textbf{B:} With a slightly higher cooperation benefit $b=3$, the system ends up with a majority of agents who cooperate with one of the 3 labels. 
In our example it is the agents with blue labels who receive the positive treatment by almost everyone, but since the labels are symmetrically defined, the dominant strategy could just as well have been only cooperating with red or with green.
We call these the $C(1)$ strategies, because they single out one of the labels as the only receiver of cooperation.
Looking at the time-series we see that all 3 $C(1)$ strategies are initially expanding by out-competing the $C(none)$ strategy.
However, when there are no more complete defectors to displace one of the $C(1)$ strategies ends up suppressing the others and eventually dominates the entire population.

\begin{figure*}
    \centering
    \includegraphics{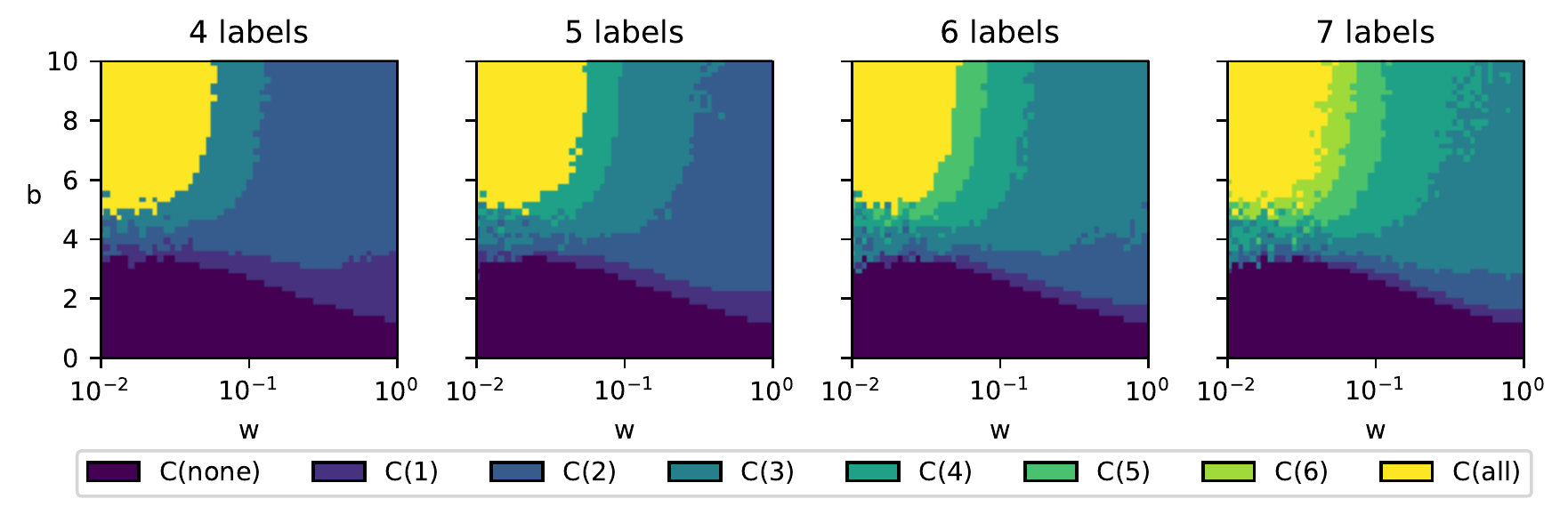}
    \caption{Phase diagrams showing the most common strategy types for squared lattices with periodic boundary conditions, 10000 agents and up to 7 labels over selection pressure \textit{w} and benefit \textit{b} with constant cost $c=1$ and constant mutation rate $\mu=0.001$. ($2.5\times10^8$ time steps for every data point)}
    \label{fig:paramscans_L>3}
\end{figure*}

\textbf{C:} The dynamics at high cooperation benefit, $b=8$, are very similar to those described in example (B), except that here it is the $C(2)$ strategies which are dominating the system.
These are the strategies which cooperate with two of the label, e.g. cooperate with red and blue neighbours, but not those with green labels.
Again we see that the system ends up being dominated by just one of these strategies.
It is worth noting that in the very early rounds, while most of the agents still have the strategy not to cooperate with anyone, there is a brief rise of agents cooperating with everyone, $C(all)$.

\textbf{D:} If, compared to example (C), the selection pressure is a little smaller, $w=0.06$, then the balance switches between the strategies $C(2)$ and $C(all)$. Here the system ends up in a state where the large majority of agents are cooperating with all of their neighbours, independently of their label.

\textbf{E:} If, however, the selection pressure is increased to a high value, $w=1$, the system fractures into mix of all strategies of strategy classes 1 and 2, with strategy class 1 being more dominant.
Interestingly it is the $C(2)$ strategies which first very quickly displace the initial defectors, but in the long run they are suppressed by the $C(1)$ strategies.

The left panel of figure \ref{fig:paramscan_L=3} is a phase diagram showing which strategy class is the most dominating as a function the cooperation benefit $b$ and the selection pressure $w$. 
One can clearly distinguish four of the five phases exemplified in figure \ref{fig:examples}.
While the distinction between (B) and (E) were clear in the examples, it cannot be seen by looking at this measure alone.

The two panels in the middle show parameter scans made at fixed cooperation benefits, $b=8.0$ and $b=2.0$ respectively, as indicated by the horizontal lines in the phase diagram.
These show what fractions of the population belong to each strategy class as a function of the selection pressure $w$.
At $b=8$ we observe two transitions between different phases:
One is a sharp transition at $w\approx 0.05$ from the phase where everyone cooperates with everyone else to the phase where almost all agents has a strategy of class 2.
The other is a smooth transition at $w\approx 0.15$ to the state where strategy class 1 and 2 coexist with class 1 being more common.
For $b=2$ we see just one sharp transition from the phase where nearly nobody cooperates with anyone to a phase strongly dominated by strategy class 1.

The two panels on the right show parameter scans similar to those in the middle, but with varying cooperation benefit $b$ and fixed selection pressures, $w=0.1$ and $w=1.0$, as marked by the vertical lines in the phase diagram. 
At $w=0.1$ we observe three different phases separated by two sharp transition.
The transition between almost no cooperation and the phase dominated by strategy class 1 happens at $b\approx 2.8$.
The phases dominated by strategy classes 1 and 2, respectively, are separated by a transition at $b\approx 4.5$ which is almost as sharp. 
At $w=1.0$ there is a sharp transition at $b\approx 1.3$ between the phase with almost no cooperation to one where almost all agents have a strategy from class 1.
When the cooperation benefit $b$ is increased above this transition, we observe a smooth change where an increasing fraction of the agents ends up with strategies from class 2, resulting in the `mixed phase' as illustrated by example E in figure~\ref{fig:examples}.

Based on these observations we can say that for a population with 3 different labels and agents arranged on a squared lattice there are sharp transitions between the phases represented by the examples A-D in figure~\ref{fig:examples} and smooth transition from B to E and from C to E.

\begin{figure}
    \centering
    \includegraphics{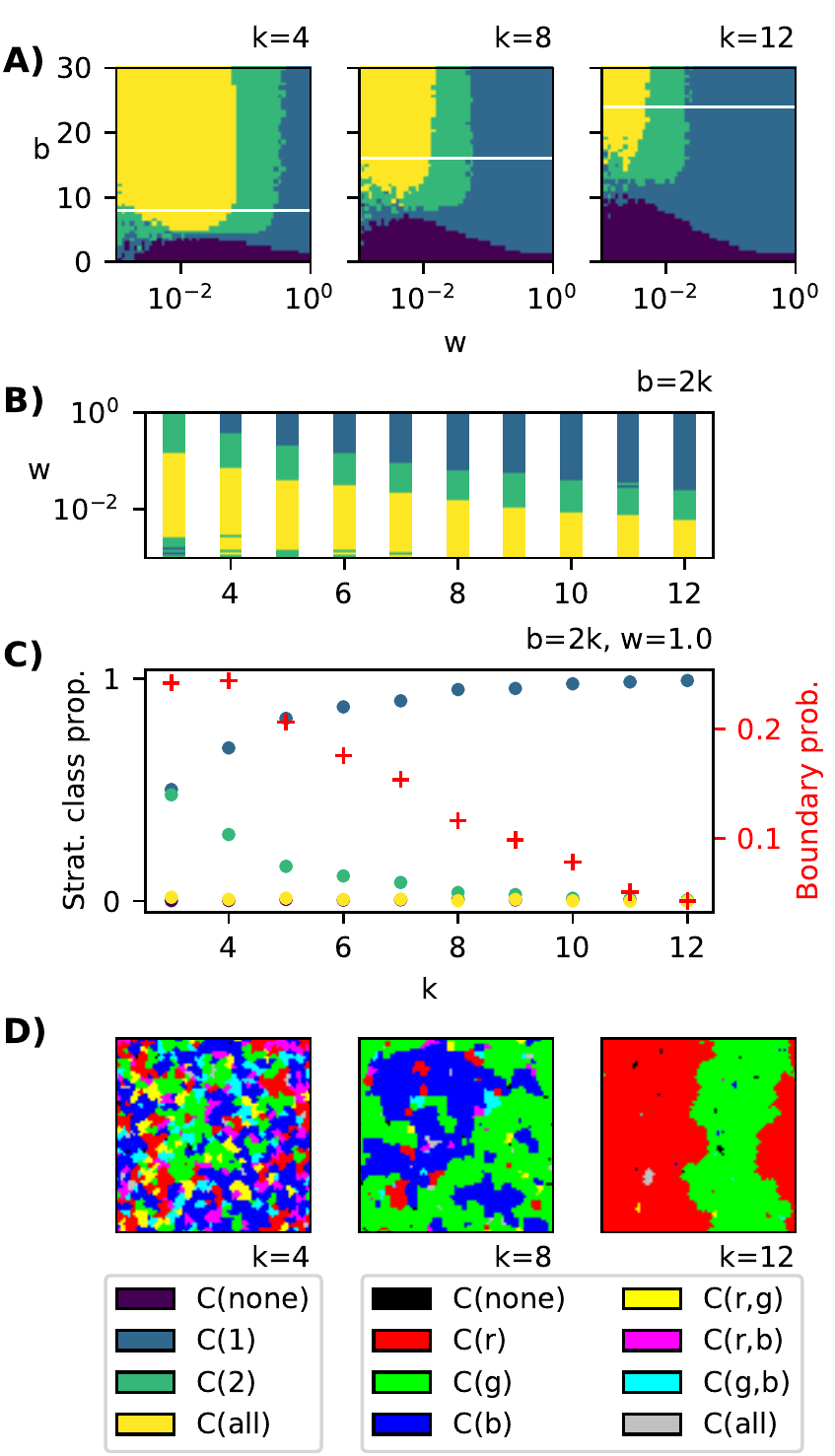}
    \caption{Results for squared lattices with periodic boundary conditions, 10000 agents and 3 labels, a constant mutation rate $\mu=0.001$ and a constant cost $c=1$ after $10^8$ time steps.
    \textbf{A:} 2D parameter scan showing the most common strategy class over benefit $b$ and selection pressure $w$ for connectivity $k=$ 4, 8 and 12.
    \textbf{B:} Parameter scans showing the most common strategy class over selection pressure $w$ and connectivity $k$ for benefit $b=2k$. 
    \textbf{C:} Strategy class proportion and the boundary probability of linked agents sharing the same strategy over the number of neighbours $k$ for benefit $b=2k$ and selection pressure $w=1$. 
    \textbf{D:} Strategy distributions for connectivity $k=$ 4, 8 and 12, benefit $b=2k$ and selection pressure $w=1$.}
    \label{fig:connectivity}
\end{figure}

\subsection{More labels}
To learn how the model behaves in systems with $L>3$ labels, figure~\ref{fig:paramscans_L>3} shows parameter scans for systems with up to 7 different labels. 
The plots show the most common strategy class as a function of cooperation benefit $b$ and selection pressure $w$. 
We see that the regions of parameter space in which the system is dominated by unconditional cooperation or defection are almost unaffected when changing the number of labels.
As we increase the number of labels, the number of strategy classes goes up as well.
Consequently, the discriminating parameter region is subdivided into an increasing number of sub-regions, characterised by which strategy-class ends up dominating the system.
This division follows a very simple structure.
Nearest to the region dominated by unconditional defection, the system will be dominated by a strategy which cooperates with one of the labels (e.g. cooperate only with blue).
When going a little further, the dominating strategies are those who cooperate with two of the labels (e.g. cooperate with blue and green, but not with the rest), and so on.
Nearest to the region of unconditional cooperation we find the system dominated by strategies which discriminate negatively against a single label (e.g. cooperate with all neighbours except the blue).

The phase diagrams in figure~\ref{fig:paramscans_L>3} only show which strategy class ends up as the most abundant at a given parameter-set.
Therefore it hides the finer structures, such as whether the system ends up being dominated by a single representative of the strategy class, or if the stationary state is a mixture of many strategies as, for example, in example E in figure~\ref{fig:examples}.
In the $L=3$ case, one can get a good sense of this transition by looking at the 1D parameter scans shown in figure~\ref{fig:paramscan_L=3} and paying attention to how big a fraction of the agents follow the most abundant strategy-class.
Similar figures, but with up to $L=7$ different labels can be found in the supplementary material.
They indicate that the parameter-region characterised by fractured coexistence (such similar to example E) doesn't depend on the number of labels.

In summary, the phase-diagram capturing the model's long-term behaviour seems to be remarkably independent of the number of labels.

\subsection{Further analysis}
The results presented so far have all been limited to the case where agents have 4 neighbours.
In this next section we will characterise how the system changes when we change the connectivity.
%That's why we now want to know what happens if we raise the connectivity.
Figure~\ref{fig:connectivity} shows simulation results from systems with 3 labels and agents arranged in grids with 3 labels and up to 12 neighbours.
Interested readers can find a detailed description of the lattice structures in the supplementary material section B.

Panel (A) shows 3 phase diagrams -- similar to those presented in the previous figures -- with 4, 8 and 12 neighbours respectively. 
It can be seen that the cooperation benefit $b$ have to be kept proportional to the connectivity $k$ to get similar results. 
This was expected based on the results from Ohtsuki et al. \cite{Ohtsuki2006} who said that cooperation occurs, if benefit to cost ratio exceeds the connectivity. It can also be seen that the system's stationary state is almost independent of cooperation benefits, when this is above $\sim 2k$ (marked by the horizontal lines). 
This means that when the cooperation benefit is high, the transition point between phases of unconditional cooperation and discrimination is almost only controlled by the selection pressure and the connectivity.
It is clear to see, that when the connectivity i higher, the transition into a discriminating phase happens at lower selection pressure.
A more detailed visualisation of this can be seen in panel (B).
Here we compare parameter scans at varying the selection pressure $w$ for connectivities ranging from $k=3$ to $k=12$ and fixed cooperation benefit $b=2k$. 
It can be seen how reduction of the selection pressure needed to increase discrimination has a slightly convex course.

Panel (C) gives some more detail about how the strategy distribution depends on connectivity at $k$ when both the selection pressure and cooperation benefit are high ($w=1$ and $b=2k$).
The round dots show what fraction of the population ends up in each strategy class, as measured on the left axis.
It can be seen that when the connectivity is higher, the gap between the strategy classes 1 and 2 becomes bigger. 
When $k\gtrsim 10$, almost all agents end up with a $C(1)$ strategy, even though all other parameters are the same as in example E in figure~\ref{fig:examples} where the stationary distribution was a disorderly mixture of small patches with different strategies.
To further quantify this difference, the red crosses show the fraction of neighbours who have different strategies (boundary fraction). 
This is measured as the number of links connecting agents with different strategies, divided by the total number of links.
The `boundary fraction' is a lot smaller when the connectivity is higher, as the system goes towards a state dominated by a single strategy.
This indicates that increasing the connectivity pushes the smooth transition between the phases represented by example E and B in figure \ref{fig:paramscan_L=3} towards higher selection pressure.
To give a more intuitive visualisation of this difference, panel (D) shows three snapshots of how the stationary strategy distribution look for three different connectivities -- $k=4$, $k=8$ and $k=12$ respectively. 

\begin{figure}
    \includegraphics[width=\linewidth]{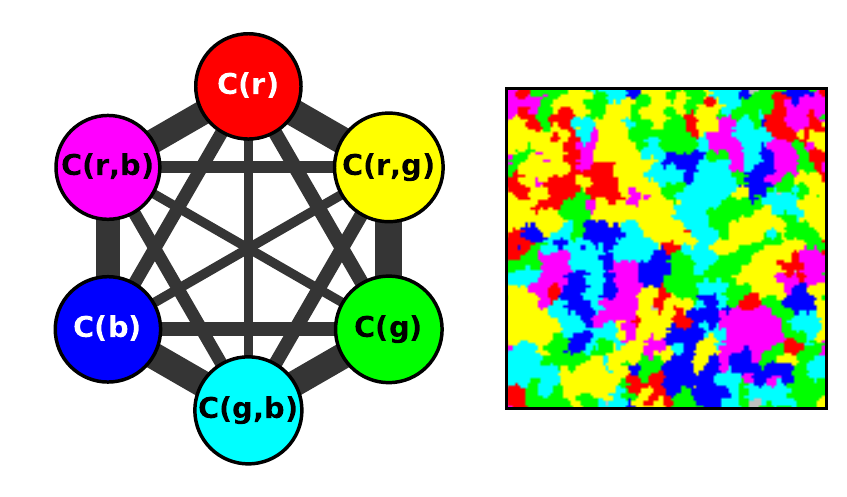}
    \caption{
    \textit{Left panel:} Graphs representation visualising the strategy distributions.
    Each node represents one of the discriminating strategies.
    The thickness of a link between two nodes is proportional to the fraction of neighbour-pairs with one agent following each of the corresponding strategies after $10^8$ timesteps averaged over 100 simulations with the following parameters:
    Each agent is connected to the $k=4$ nearest neighbours in a $100\times100$ square lattice, and has one of $L=3$ possible labels chosen independently with uniform probability.
    We used cooperating benefit $b=8$, selection pressure $w=10^{-0.54}$, and mutation rate $\mu=0.001$, and we initiated all simulations with all agents following the `defect all' strategy.
    \textit{Right panel:} Example of one of the 100 datapoints combined to make the graph.
    }
    \label{fig:neighbourSimilarity}
\end{figure}

It is not immediately obvious, but a careful study of snapshots like the one in figure~\ref{fig:connectivity} with connectivity $k=4$ or example E in figure~\ref{fig:examples} reveals an interesting pattern in how frequently different strategies occupy neighbouring nodes.
Patches of a $C(2)$ strategy -- which cooperating with two of the labels -- are less likely to be found next to patches of the $C(1)$ strategy which cooperates with the third label than next to the two other $C(1)$ strategies.
E.g. yellow patches (cooperate with green and red) share, on average, more border with the green or red patches than with the blue.
In order to quantify this observation we can draw graphs like the one in figure~\ref{fig:neighbourSimilarity}, where each node represents one of the discriminating strategies, and the thickness of the link between two nodes is proportional to the number of neighbour-pairs with one neighbour following each of the corresponding strategies (on average over an ensemble of individual simulations).
The result varies with the choice of parameters.
Here we have chosen to emphasise an example at cooperation benefit $b=8$ and selection pressure $w=10^{-0.54}$. 
This point is near the transition between the phases dominated by $C(1)$ and $C(2)$ strategies, so all the discriminating strategies are approximately equally abundant.
We have chosen not to include the unconditional strategies, since they are almost non-existent at these parameters.
We have also chosen not to visualise the self-links, because these would be many times stronger than the links between different strategies.
The figure shows that strategies are more likely to occupy neighbouring when they agree on how to behave towards more of the labels.
A handful of other examples are included in the supplementary material section C.
These show that the result is qualitatively robust at for a wide range of parameters near the transition between phases where different strategy-classes coexist, except for very low values of selection pressure where the signal drowns in noise. 

While all the results presented in the so far have been obtained from systems where the agents are arranged in regular lattices, it is straightforward to apply our model to other topologies.
In the supplementary material section A we have included figures showing how the model behaves on both random regular graphs and Erd\H{o}s-R\'{e}nyi (binomial) random graphs.
The main results are very similar to those obtained on regular graphs, but there are a few differences worth mentioning.
One difference is, that without the low-dimensional spatial grid structure, the model has an even stronger tendency to end with almost all agents applying the same strategy.
This results in some expanded bistable parameter regions around the transitions  between the different phases.
In these regimes, the systems will end up being dominated by one single strategy, but it is unpredictable from which strategy class.
Another difference is caused by the variation of connectivities in the Erd\H{o}s-R\'{e}nyi graphs.
Agents with more neighbours can potentially end up with higher payoffs. 
Since the model has a non-linear tendency to select `the richest' neighbours, 
the dynamics tend to being dominated by the local structures around highly connected agents.
This can lead to scattered strategy distributions when the selection pressure is high, $w\gtrsim0.1$ .

\section{Discussion}
\label{discussion}
In this paper we have studied a spatial prisoner's dilemma model in which agents marked with 
meaningless labels imitate the strategies of their neighbours -- preferably the richest ones. 
It is a well know result discovered by Ohtsuki et al. \cite{Ohtsuki2006}, that in the special case 
with only $L=1$ label, i.e.\ when all agents are identical, cooperation is the evolutionarily 
stable strategy when the cooperation benefit is greater than the average connectivity, i.e. $b>k$.
Our simulations confirm that this simple rule also applies to systems with a multitude of different labels, 
as long as the selection pressure is sufficiently low.
We have also confirmed that states of persistent hierarchical discrimination can emerge through 
spontaneous symmetry breaking, which was previously only demonstrated in the case of binary labels 
($L=2$) \cite{jensen2019imitating}.
We have investigated systems with up to $L=7$ different labels, and we found that the transition 
points between unconditional and discriminating strategies are essentially independent of the number of labels.
This quantitative robustness is interesting when considering that the two unconditional strategies 
(cooperate/defect against all) constitute a rapidly decreasing fraction of the number of possible strategies, 
which increases exponentially with the number of labels.

While the main results known from the cases with $L=1$ and $L=2$ are essentially unchanged in systems with $L>2$, 
our investigations have also revealed some new phenomena. One finding is that the parameter-region dominated 
by discriminating strategies is subdivided into distinct phases dominated by different strategy-classes.
While one cannot know from the onset of a simulation which strategies are going to dominate in the system, 
it is possible to make reliable predictions about how many of the labels these strategies treat with cooperation and defection.
This is similar to two discriminating phases ``cooperate with the majority'' and ``cooperate with the minority'' 
which were detected in the binary-label model.

Another fascinating phenomenon which can only be observed when $L>2$ is that neighbours are 
more likely to have similar strategies. Since strategies are copied from neighbour to neighbour, 
it is no surprise that the model forms patches of agents agreeing on a single strategy. %(just think about the voter model [REF])
There are, however, no mechanisms in our model which explicitly favour neighbours with similar strategies.
It is therefore surprising to observe that it is more likely for neighbours to have more similar strategies, 
when we measure similarity between two strategies as the number of label towards which they agree on what action 
to take. We do not yet understand this observation and finding an explanation will require further research.

The model described in this paper is not the only one in the literature of evolutionary game theory 
which is designed to investigate the combined effect of spatial structure and distinguishable agents.
One example is the work by Garcia et al.\ \cite{Garcia2014} who demonstrated that introducing tags 
may reduce the amount of cooperation in a structured population through a mechanism they have called 
``the evil green-beard effect''. The outcome that introducing tags can lower the total amount of 
cooperation through negative discrimination is also present in the model presented in this paper,
yet the dynamics leading to this effect is quite different. Our model tends to end up in in a 
stationary state, whereas theirs exhibits cyclic behaviour which is characteristic for tag-based cooperation 
\cite{Riolo2001, traulsen2003, Traulsen2007, Fu2012, Cohen2012, Hadzibeganovic2014,Garcia2014}.
One of the main differences is that the models of tag-based cooperation treat the tags as a part of 
the variable state subject to the evolution dynamics. Our labels, on the other hand, are immutable 
properties of the agents, determined at the beginning of the simulation. This can be seen as an 
approximation of a system in which behaviours adapt on a much faster timescale than physical appearances
as, for example, if the dynamics are thought of as a type of learning via social imitation, 
while the labels represent genetically determined physical traits.
This is reminiscent of the close connection between multi-agent reinforcement learning and 
evolutionary dynamics \cite{borgers1997, nicole2017stochastic}.
Another important assumption in our model is that agents choose which of their neighbours to imitate 
solely based on their fitness. In particular it doesn't consider the labels -- neither its own nor those 
of the neighbours -- or how they interact with the imitated strategy. This can be interpreted as if 
the agents are not aware of their own labels.
It is difficult to imagine, that humans should carry easily observable markers without being aware of them themselves.
Our softer interpretation is that the strategies represent subconscious biases exempted from rational reasoning.
Experimental evidence suggesting that humans do exhibit ingroup devaluation (or outgroup favouritism) 
\cite{Rudman2002,Jost2002,March2014} supports this viewpoint and finds a possible representation in the model. 

In conclusion, the model we investigated in this paper demonstrates that imitating successful behaviours 
may lead to the emergence of persistent hierarchical discrimination in a population where agents are marked 
with observable, but otherwise meaningless, labels. We found this to be a remarkably robust phenomenon 
with respect to the number of labels. A central mechanism of the emergence of hierarchical social 
structures in the model is spontaneous symmetry breaking, transforming initial randomness into persistent fates.  

\section{\label{ref}References}
\bibliography{biblo.bib}

\end{document}

\preprint{APS/123-QED}

\title{Supplementary material to: Discrimination emerging through spontaneous symmetry-breaking in a spatial prisoner's dilemma model with multiple labels}

\author{Gorm Gruner Jensen}
\email{ggjensen@itp.uni-bremen.de}
\author{Frederik Tischel}
\author{Stefan Bornholdt}
\affiliation{%
Institute for Theoretical Physics, University of Bremen, 28359 Bremen, Germany
}

\date{\today}

\maketitle{}

\section*{Supplement A: Random graphs}
\setcounter{figure}{0} 
\renewcommand\thefigure{A\arabic{figure}}
All the figures presented in the main text are based on systems where agents are arranged in regular 2D-lattices. 
To investigate to what extend the results depend on this topology, we have reproduced some of the results using both random regular graphs and Erd\H{o}s-R\'{e}nyi graphs. 

When we use random regular graphs we generally find a very similar results to those obtained on regular lattices.
However, there is one difference worth pointing out.
Since the random regular graph has very small diameter and cluster-coefficient, it is not possible to form `patches' in the same way as on a 2D lattice.
As a result the entire system is even more likely to end up being dominated by a single strategy.
The amplification of the `winner takes all' dynamics also makes it less predictable to which strategy-class the winning strategy belongs, as an early victory becomes more important.
Therefore, the system expresses more bistable behaviour at the parameters near the transitions between the different phases.

On Erd\H{o}s-R\'{e}nyi graphs the results are in addition a lot more fuzzy at high selection pressures $w \gtrsim 0.1$.
An agent with higher connectivity has the potential to get a higher profit and consequently much higher fitness.
Because of this, the model is not very robust to local variations in connectivity.

\subsection*{Phase diagram}
\begin{figure*}[!ht]
    \centering
    \includegraphics{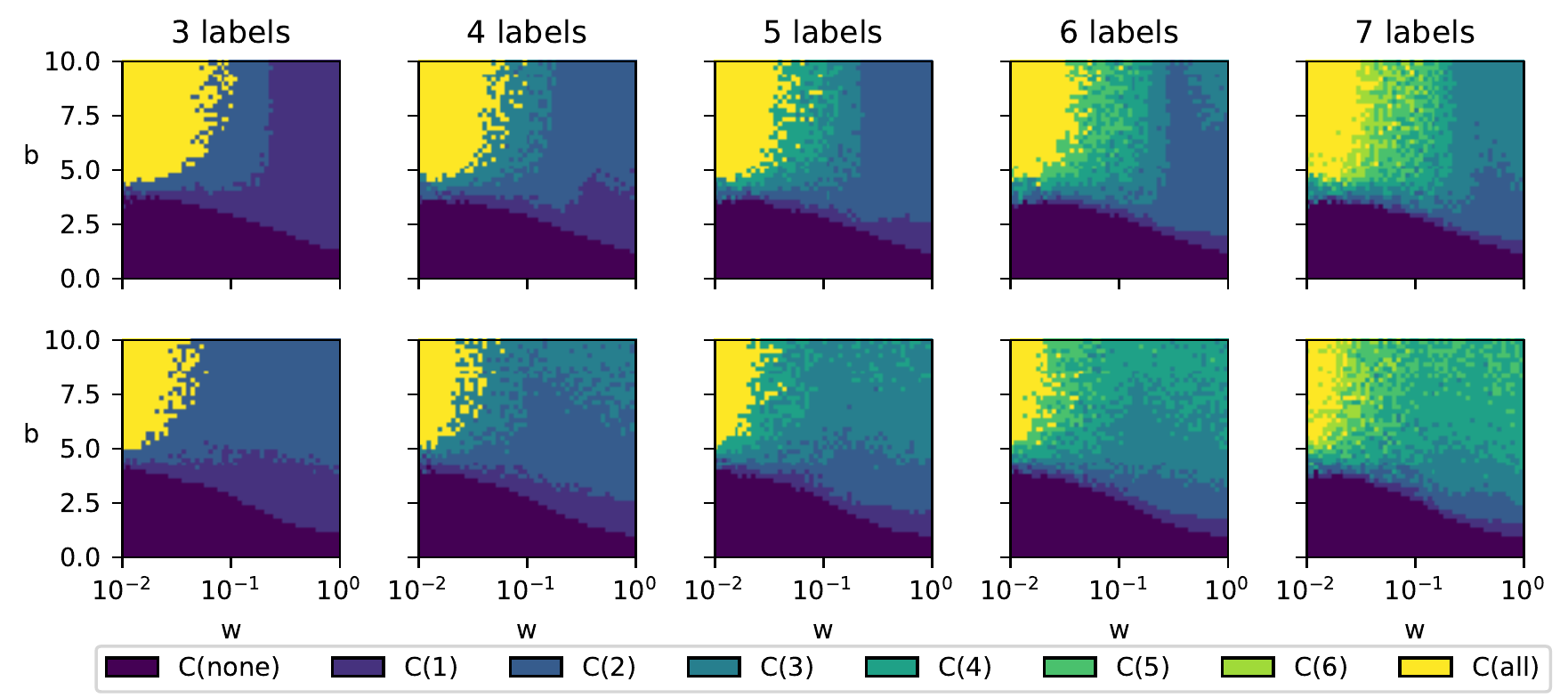}
    \caption{
    Phase diagrams showing the most common strategy type for Random Regular Graphs(top) with connectivity 4 and Erd\H{o}s-R\'{e}nyi Random Graph(bottom) with avg connectivity 4, 10000 agents and up to 7 labels over selection pressure \textit{w} and benefit \textit{b} with constant cost $c=1$ and constant mutation rate $\mu=0.001$. ($2.5\times10^8$ timesteps for every data point). Essentially the same as figure~3, but using different topologies.
    }
    \label{fig:A1}
\end{figure*}

\clearpage

\subsection*{1D parameter scan with constant cooperation benefit $b$}
\begin{figure*}[!ht]
    \centering
    \includegraphics{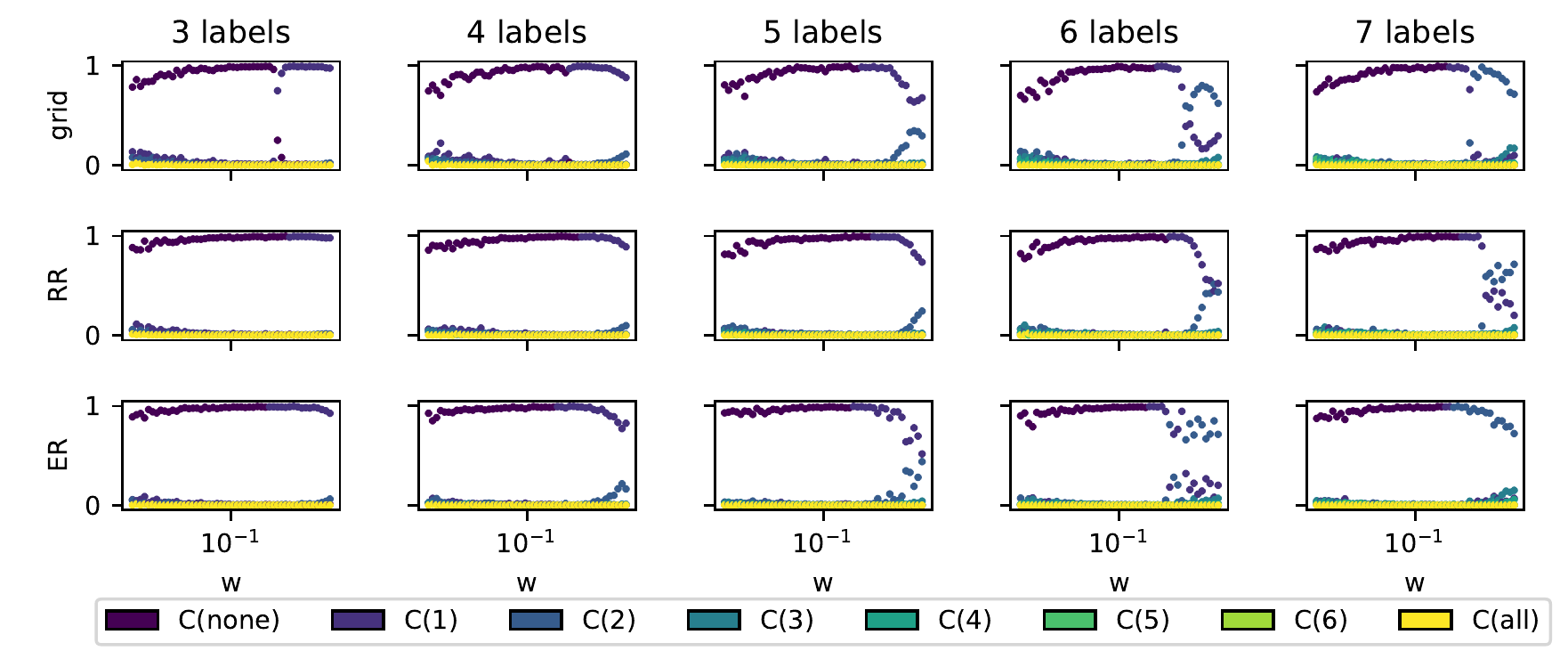}
    \caption{
    Parameter scans showing strategy class proportion over selection pressure \textit{w} for Square Lattices(grid), Random Regular Graphs(RR) and Erd\H{o}s-R\'{e}nyi Random Graphs(ER) with fixed benefit \textit{b=2} and up to 7 labels. ($2.5\times10^8$ timesteps for every data point)
    }
    \label{fig:A2}
\end{figure*}

\begin{figure*}[!ht]
    \centering
    \includegraphics{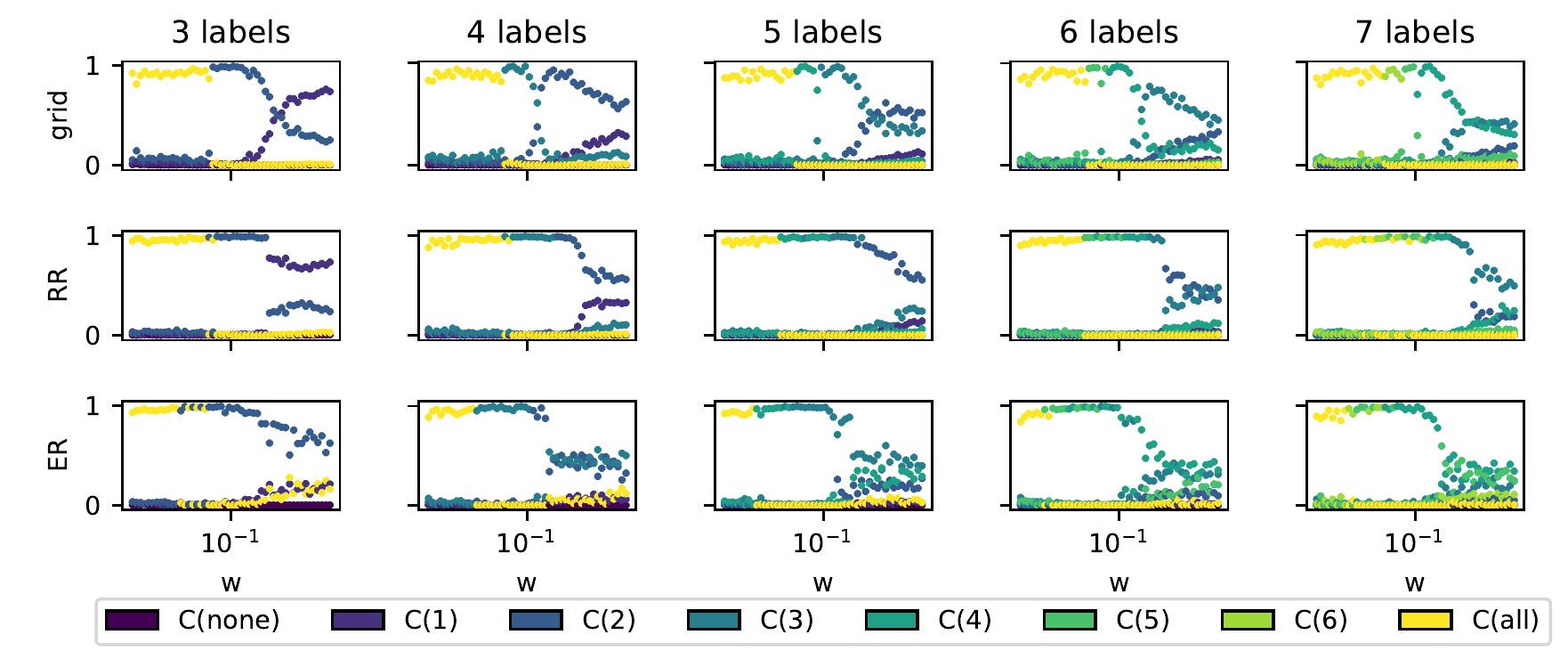}
    \caption{
    Parameter scans showing strategy class proportion over selection pressure \textit{w} for Square Lattices(grid), Random Regular Graphs(RR) and Erd\H{o}s-R\'{e}nyi Random Graphs(ER) with fixed benefit \textit{b=8} and up to 7 labels. ($2.5\times10^8$ timesteps for every data point)
    }
    \label{fig:A3}
\end{figure*}

\clearpage

\subsection*{1D parameter scan with constant selection pressure $w$}
\begin{figure*}[!ht]
    \centering
    \includegraphics{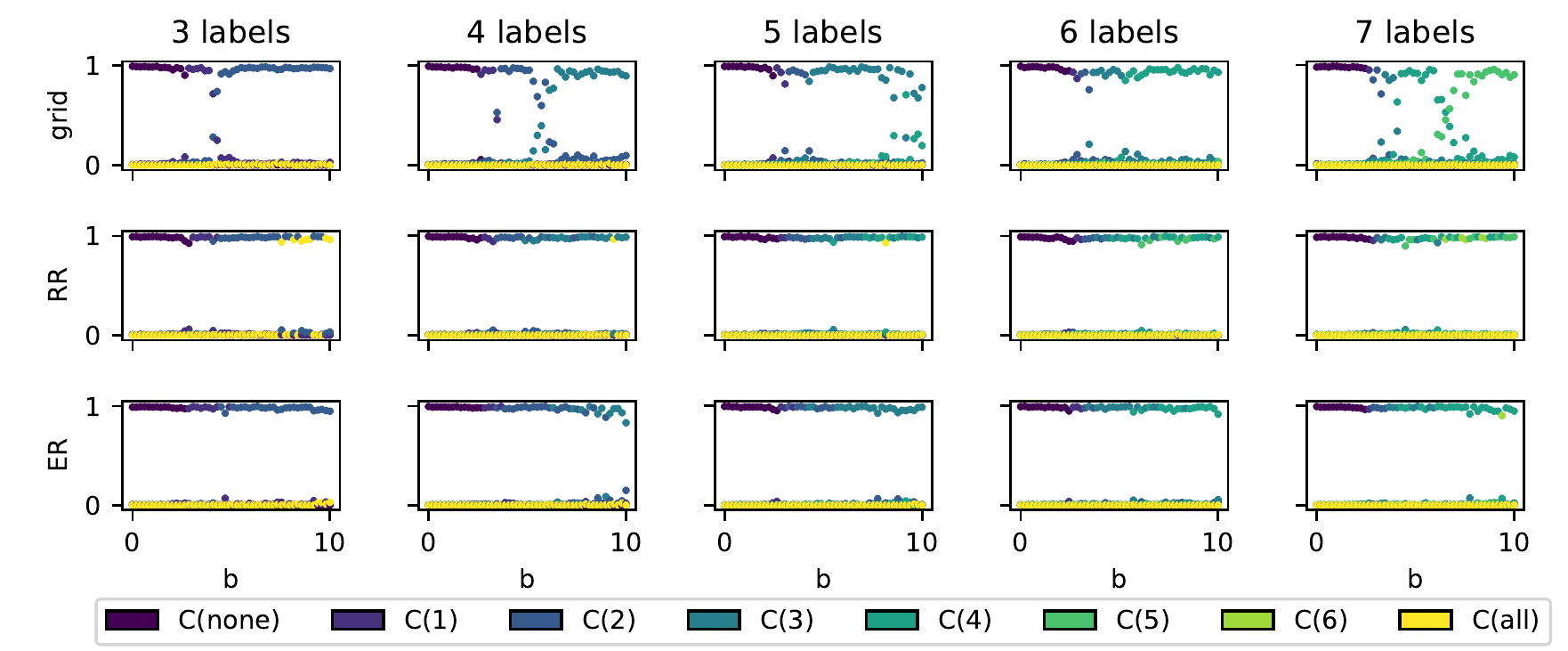}
    \caption{
    Parameter scans showing strategy class proportion over benefit \textit{b} for Square Lattices(grid), Random Regular Graphs(RR) and Erd\H{o}s-R\'{e}nyi Random Graphs(ER) with fixed selection pressure \textit{w=0.1} and up to 7 labels. ($2.5\times10^8$ timesteps for every data point)
    }
    \label{fig:A4}
\end{figure*}

\begin{figure*}[!ht]
    \centering
    \includegraphics{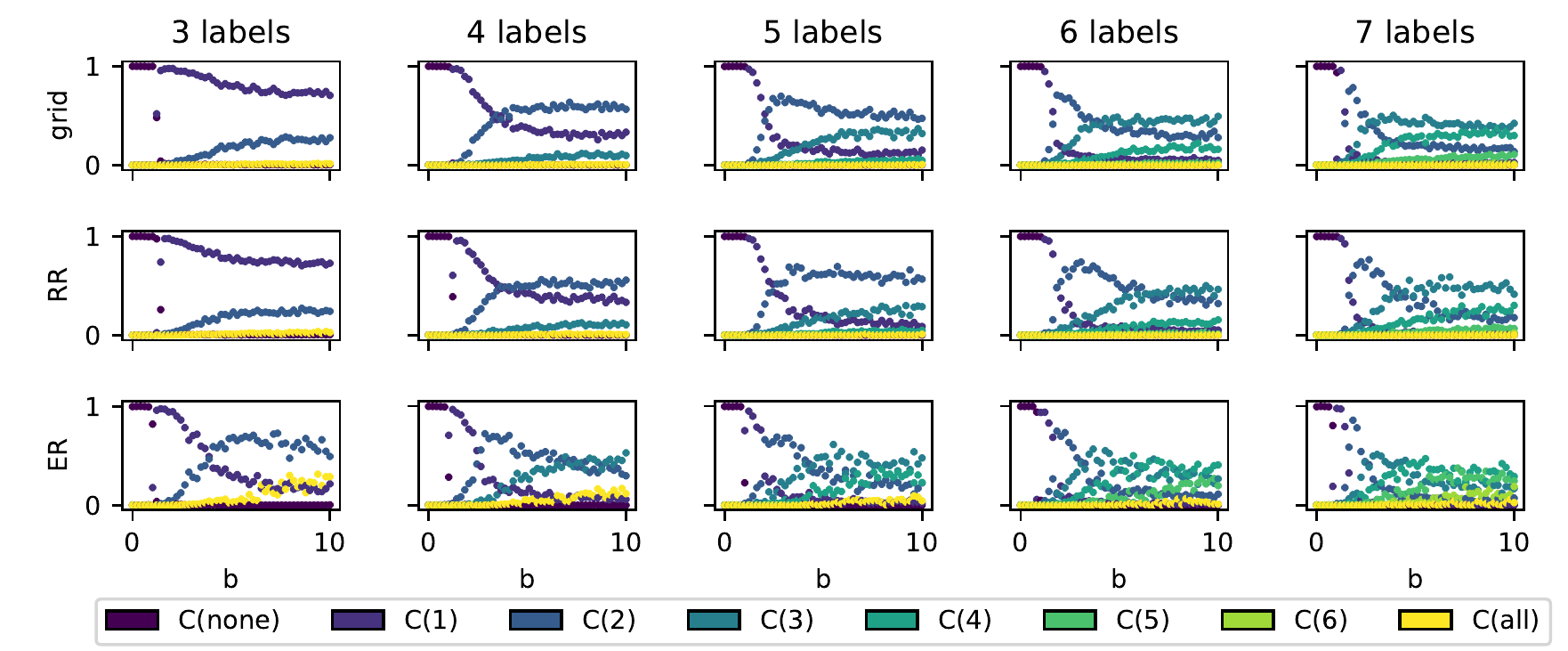}
    \caption{
    Parameter scans showing strategy class proportion over benefit \textit{b} for Square Lattices(grid), Random Regular Graphs(RR) and Erd\H{o}s-R\'{e}nyi Random Graphs(ER) with fixed selection pressure \textit{w=1.0} and up to 7 labels. ($2.5\times10^8$ timesteps for every data point)
    }
    \label{fig:A5}
\end{figure*}

\clearpage

\subsection*{Varying connectivity}
\begin{figure*}[!ht]
    \centering
    \includegraphics{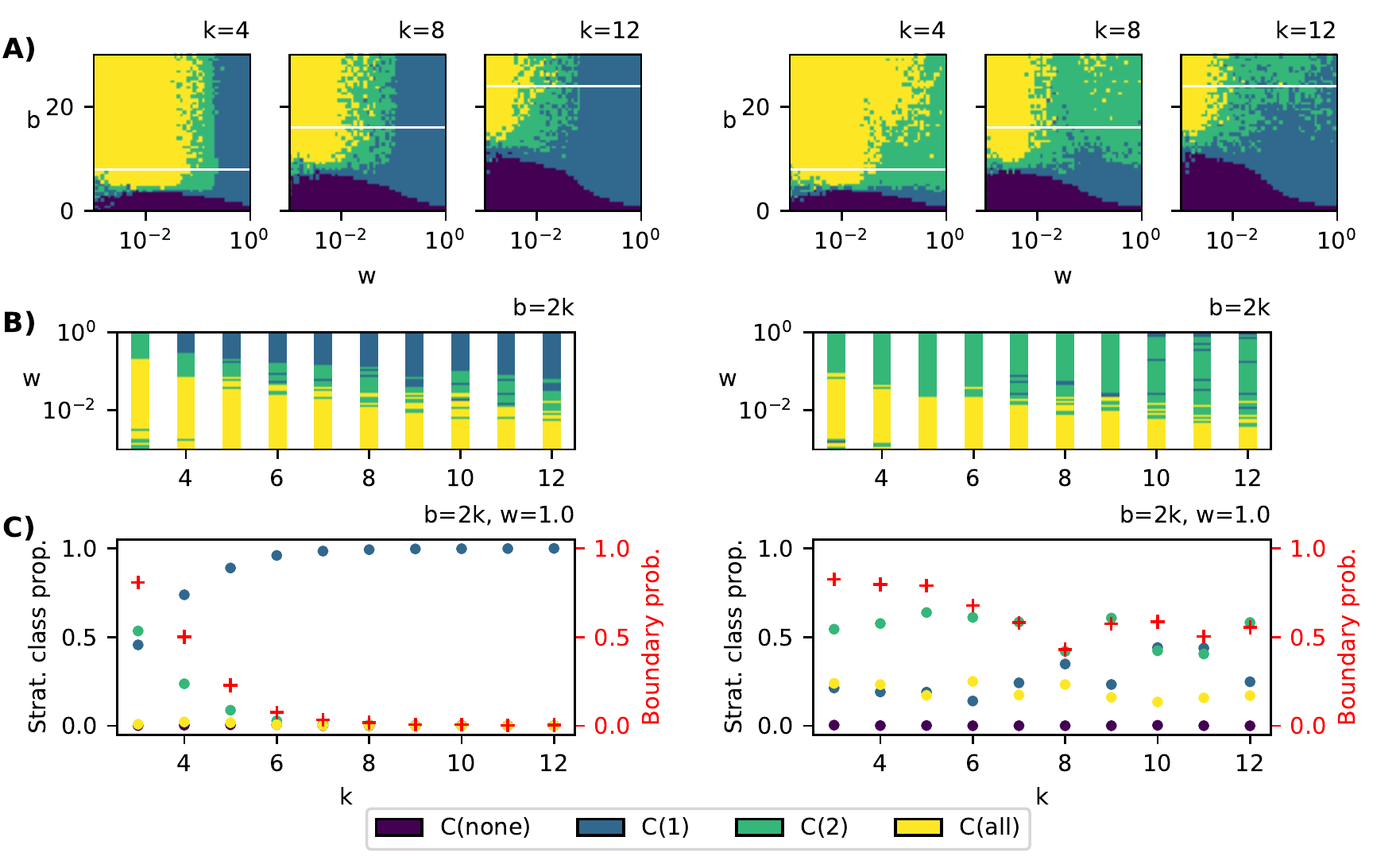}
    \caption{
    Results for Random Regular Graphs(left) and Erd\H{o}s-R\'{e}nyi Random Graphs(right), with 10000 agents and 3 labels, a constant mutation rate $\mu=0.001$ and a constant cost $c=1$ after $10^8$ time steps.\textbf{A:} 2D parameter scan showing the most common strategy class over benefit $b$ and selection pressure $w$ for connectivity $k=$ 4, 8 and 12.\textbf{B:} Parameter scans showing the most common strategy class over selection pressure $w$ and connectivity $k$ for benefit $b=2k$. \textbf{C:} Strategy class proportion and the boundary probability of linked agents sharing the same strategy over the number of neighbours $k$ for benefit $b=2k$ and selection pressure $w=1$. Essentially the same as figure~4 in the main text, but with different topologies.
    }
    \label{fig:A6}
\end{figure*}

\clearpage

\section*{Supplement B: Lattice neighbourhoods}
\setcounter{figure}{0} 
\renewcommand\thefigure{B\arabic{figure}}
\label{appendix:neighbourhood}
In figure~4 in the main text, we have presented results based on systems where agents are distributed on lattices with varying neighbourhood sizes.
While there are standard neighbourhoods of size $4$, $6$, $8$, and $12$ it is more ambiguous how to design a lattice where all agents have for example an odd number of neighbours.
Here we present a visualisation of which neighbourhoods we have used in our simulations.
The choices are not unique, but we see no reason to believe that small variations should have a qualitative impact on the results.
\begin{figure}[!ht]
    \centering
    \includegraphics{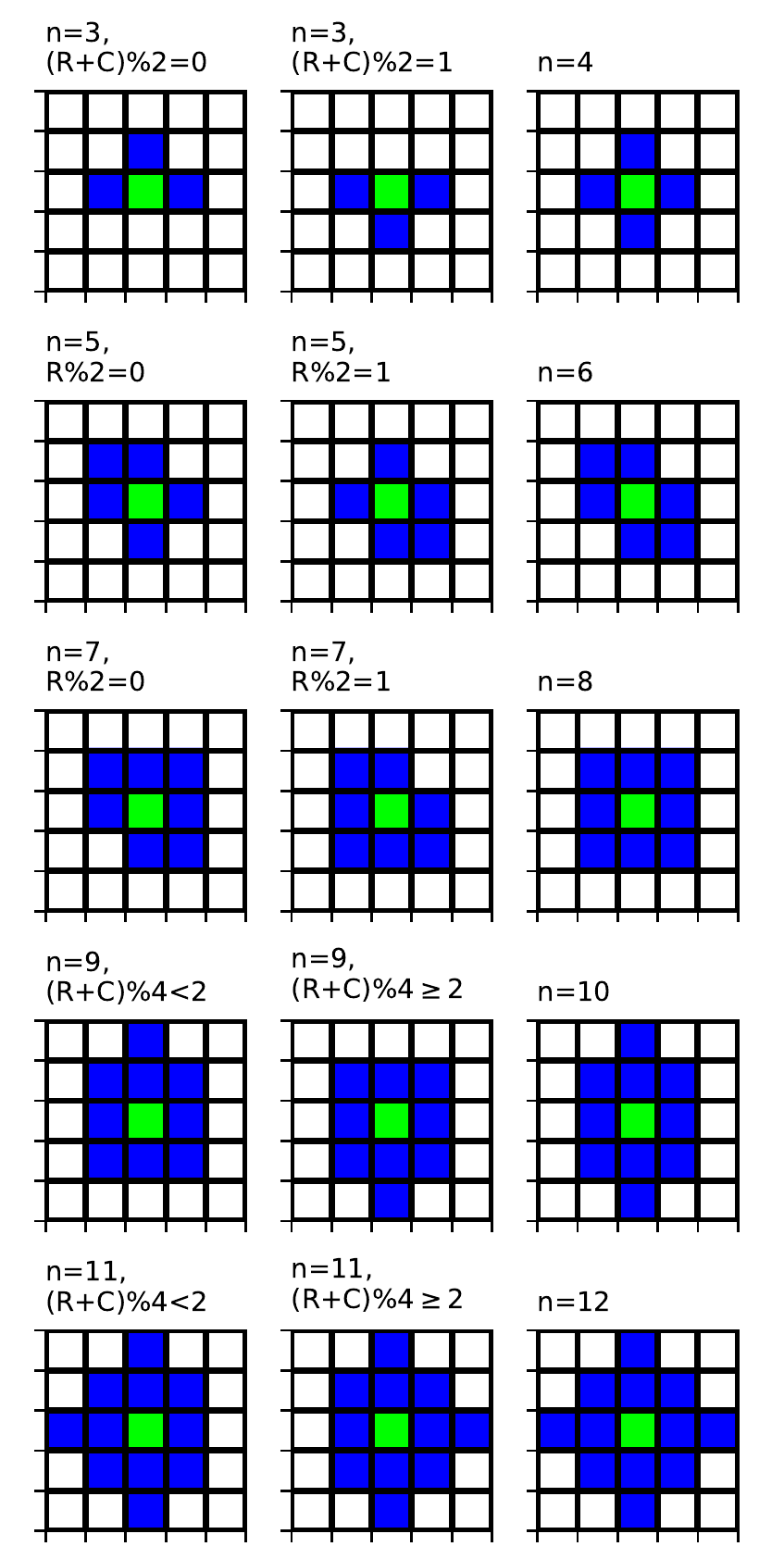}
    \caption{Neighbourhood(blue) of an agent(green) on a squared lattice with connectivity n. Row(R) and Column(C)}
    \label{fig:B1}
\end{figure}

\clearpage

\section*{Supplement C: Neighbours have similar strategies}
\setcounter{figure}{0} 
\renewcommand\thefigure{C\arabic{figure}}
\renewcommand\thetable{C\arabic{table}}
\vspace{-2mm}
Figure~5 in the main text illustrates that neighbours are more likely to have similar strategies, where the `similarity' between two strategies is measured by the number of labels towards which they agree on what action to take.
The parameters in the example were chosen such that all the different discriminating strategies are approximately equally abundant. 
Here we show a handful of other examples, to demonstrate that the phenomenon is consistent for a large variety of parameters, but that it breaks down when the selection pressure is very low.
In addition we have included a table with the numbers used to draw the graphs.
\begin{figure*}[!ht]
    \centering
    \includegraphics[width=0.8\linewidth]{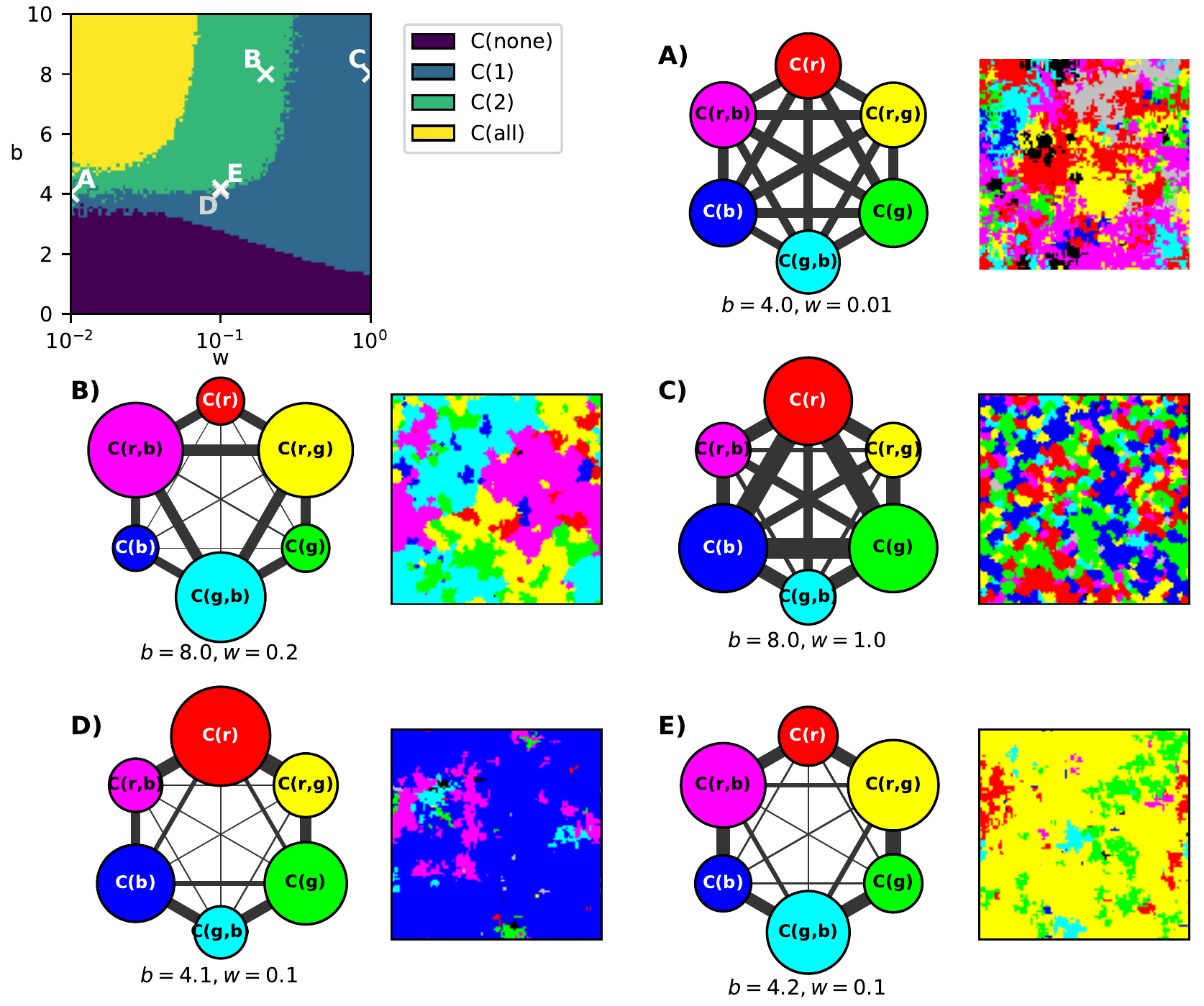}
    \caption{
    Strategy distributions visualised by graphs.
    Each node represent a strategy and the area is proportional to the fraction of agents following it.
    The thickness of a link between two nodes is proportional to the fraction of neighbour-pairs with one agent following each of the represented strategies.
    Each graph is the average results of 100 simulations at different parameters ($b$ and $w$), as marked in the phase-diagram (upper left panel).
    All the graphs are produced in systems with:
    Agents arranged in a $100\times100$-square lattice connected to their $k=4$ nearest neighbours.
    The $L=3$ different labels are chosen randomly at the beginning of each simulation with an independent uniform distribution.
    Each simulation is initiated with all agents following the "defect all" strategy, and the fraction of strategies and neighbour-pairs are measured after $10^8$ timesteps.
    The mutation rate is $\mu=0.001$.
    }
    \label{fig:C1}
\end{figure*}

\begin{table}[!h]
    \centering
    \begin{tabular}{|c||r|r||r|r||r|r||r|r||r|r||r|r|}
        \hline
        Data & \multicolumn{2}{c||}{Fig.~5} & \multicolumn{2}{c||}{A} & \multicolumn{2}{c||}{B} & \multicolumn{2}{c||}{C} & \multicolumn{2}{c||}{D} & \multicolumn{2}{c|}{E} \\
        \hline
        link-type & mean & std & mean & std & mean & std & mean & std & mean & std & mean & std \\
        \hline
        \hline
        $C(1) - C_a(2)$ & 358 & 22 & 241 & 49 & 242 & 31 & 359 & 21 & 306 & 100 & 333 & 71 \\
        $C(1) - C(1)$ & 188 & 33 & 253 & 100 & 26 & 11 & 515 & 41 & 108 & 69 & 50 & 42 \\
        $C(2) - C(2)$ & 176 & 31 & 216 & 98 & 279 & 52 & 105 & 17 & 43 & 36 & 116 & 71\\
        $C(1) - C_b(2)$ & 130 & 21 & 254 & 67 & 39 & 14 & 232 & 21 & 34 & 24 & 39 & 24\\
        \hline
    \end{tabular}
    \caption{
    The graphs above are generated on the background of 100 individual simulations each.
    After each simulation we count the occurrences of each of the four types of connections.
    The means and standard deviations each ensemble are presented in this table.
    }
    \label{tab:C1}
\end{table}